\documentclass[a4paper,english,numberwithinsect]{eurocg23}

\usepackage{microtype} % if unwanted, comment out
\usepackage[full]{complexity}
\usepackage[textsize=footnotesize]{todonotes}
\usepackage{enumerate}
\usepackage[capitalise]{cleveref}
\usepackage{amssymb,amsmath,amsfonts}
\usepackage{cite}
\usepackage{graphicx}
\usepackage[textsize=footnotesize]{todonotes}
\usepackage{booktabs}
\usepackage{array}

\graphicspath{{./Figures/}}

\newcommand{\timest}{\texttt{time}}
\newcommand{\charac}{\texttt{char}}
\newcommand{\ints}{\texttt{inter}}

\newcommand{\ILPone}{\ensuremath{\mathrm{ILP1}}}
\newcommand{\ILPtwo}{\ensuremath{\mathrm{ILP2}}}
\newcommand{\ILPoneml}{\ensuremath{\mathrm{ILP1ML}}}
\newcommand{\ILPtwoml}{\ensuremath{\mathrm{ILP2ML}}}
\newcommand{\Piplsim}{\ensuremath{\mathrm{P_s}}}
\newcommand{\Piplpat}{\ensuremath{\mathrm{P_p}}}

\newcommand{\new}[1]{\textcolor{black}{#1}}

\title{Crossing Minimization in Time Interval Storylines\footnote{Alexander Dobler was supported by the Vienna Science and Technology Fund (WWTF)  [10.47379/ICT19035]. Anaïs Villedieu was supported by the Austrian Science Fund (FWF) under grant P31119. We thank Martin Gronemann for the initial discussion.
}}
\titlerunning{Crossing Minimization in Time Interval Storylines}

\author{Alexander Dobler}
\author{Martin Nöllenburg}
\author{Daniel Stojanovic}
\author{Anaïs~Villedieu}
\author{Jules Wulms}
\affil{Algorithms and Complexity Group, TU Wien\\
  \texttt{\{adobler|noellenburg|avilledieu|jwulms\}@ac.tuwien.ac.at,\\e11907566@student.tuwien.ac.at}}
\authorrunning{A. Dobler et al.}
\ArticleNo{36}

\begin{document}

\maketitle

\begin{abstract}
  Storyline visualizations are a popular way of visualizing characters and their interactions over time: Characters are drawn as $x$-monotone curves and interactions are visualized through close proximity of the corresponding character curves in a vertical strip. Existing methods to generate storylines assume a total ordering of the interactions, although real-world data often do not contain such a total order. Instead, multiple interactions are often grouped into coarser time intervals such as years. We exploit this grouping property by introducing a new model called storylines with time intervals and present two methods to minimize the number of crossings and horizontal space usage. We then evaluate these algorithms on a small benchmark set to show their effectiveness.
\end{abstract}

\section{Introduction}
\begin{figure}[th!]
	\centering
	\includegraphics[width=\linewidth]{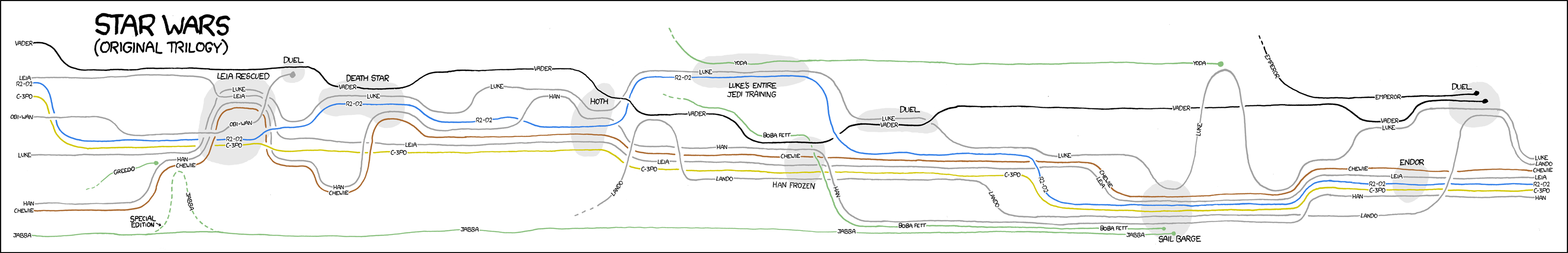}
	\caption{The xkcd comic showing a storyline of the Star Wars movie.
	}
	\label{fig:xkcd}
\end{figure}

Storyline visualizations are a popular way of visualizing characters and their interactions through time. They were popularized by Munroe's xkcd comic~\cite{munroe_movie_2009} (see~\cref{fig:xkcd} for a storyline describing a movie as a series of scenes through time, in which the characters participate). A character is drawn using an $x$-monotone curve, and the vertical ordering of the character curves varies from left to right. A scene is represented by closely gathering the curves of characters involved in said scene at the relevant spot on the $x$-axis, which represents time. Storylines attracted significant interest in visualization research, especially the question of designing automated methods to create storylines adhering to certain quality criteria~\cite{LiuWWLL13,OgawaM10,TanahashiM12}.

\begin{figure}
	\centering
	\includegraphics{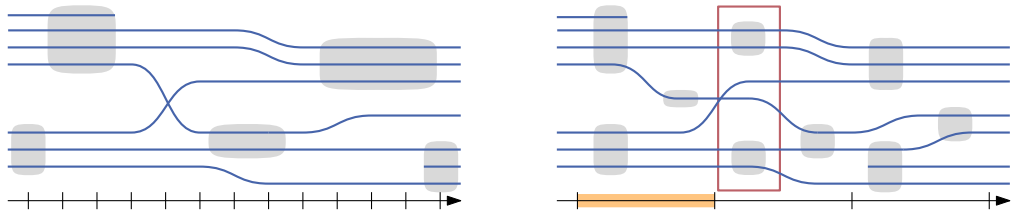}
	\caption{\textbf{\textsf{(a)}} A classic storyline with blue character lines. Interactions are shown in gray, they happen on specific timestamps and have a duration. \textbf{\textsf{(b)}} A time interval storyline. The horizontal orange segment shows a slice, every interaction on this segment has the same timestamp. A layer is highlighted in red, containing two interactions with the same timestamp but not sharing a character.
	}
	\label{fig:model}
\end{figure}

\looseness=-1
While different design optimization goals can be specified, most theoretical research has been focused on crossing minimization~\cite{GronemannJLM16,KostitsynaNP0S15} and variants like block crossing minimization~\cite{DijkFFLMRSW17,DijkLMW17}. 
This problem is \NP-hard~\cite{KostitsynaNP0S15,DijkFFLMRSW17} and is commonly solved using ILP and SAT formulations~\cite{GronemannJLM16,DijkLMW17}; \new{it has many similarities with the metro line crossing minimization problem~\cite{BekosKPS07,agm-mcp-08,fp-mcmhatc-13,bnuw-miecw-07}}. 
Recently a new model for storylines was proposed by Di Giacomo et al.~\cite{GiacomoDLMT20} that allows for one character to be part of multiple interactions at the same point in time, by modeling each character as a tree rather than a curve. Using this model, it is possible to represent data sets which have a more loosely defined ordering of interactions. 
Furthermore, authorship networks have been a popular application for storylines visualizations~\cite{GiacomoDLMT20,herrmann_2022}. In this paper we introduce \emph{time interval} storylines, an alternative approach to visualize data sets with less precise temporal attributes. In the time interval model, a set of discrete, totally ordered timestamps is given, which serve to label disjoint time intervals (e.g., the timestamp 2021 represents all interactions occurring between January and December of the year 2021). Each interval is represented in a storyline as a horizontal section in which all interactions with the same timestamp occur. The horizontal ordering within this section, however, does not correspond to a temporal ordering anymore (see \cref{fig:model}). For example, an authorship network often sorts publications by year. 
In a traditional storyline model, the complete temporal ordering of the interactions must be provided. 
\new{Previous models like the one by van Dijk et al.~\cite{DijkLMW17} can place multiple disjoint interactions in the same vertical layer, but the assignment of interactions to the totally ordered set of layers must be given as input.
Unlike the traditional model, we have no pre-specified assignment of interactions to layers, but interactions with the same timestamp can be assigned to any layer within the time interval of this timestamp.}

\subparagraph*{Problem setting.}
We are given a triple $\mathcal{S}=(\mathcal{C},\mathcal{I},T)$, of characters $\mathcal{C}=\{c_1,\dots,c_n\}$, interactions $\mathcal{I}=\{I_1,\dots,I_m\}$, and totally ordered timestamps $T=\{t_1,\dots,t_p\}$ as input. Each interaction $(C_j,t) = I_j\in \mathcal{I}$ consists of a set~$C_j\subseteq \mathcal{C}$ of characters involved in the interaction and a timestamp~$t\in T$ at which the interaction~$I_j$ occurred, respectively denoted by $\charac(I_j)=C_j$ and $\timest(I_j)=t$.
A subset of interactions can form a \emph{layer}~$\ell$, when for every pair of interactions $I,I'$ in~$\ell$, $\timest(I)=\timest(I')$. 
A time interval storyline is composed of a sequence of layers to which interactions are assigned. Intuitively, a layer represents a column in the storyline visualization, in which interactions are represented as vertical stacks. Thus, to each layer we associate a vertical ordering of~$\mathcal{C}$.
Consider the set $S$ containing all interactions with timestamp $t$, we call the union of layers containing $S$ a \emph{slice}.

Characters are represented with curves passing through each layer at most once. To represent an interaction $I=(C,t)$ in a layer $\ell$, the ordering of the characters in $\ell$ must be such that the characters of $C$ appear consecutively in that ordering.
For a pair~$I, I'$ of interactions in the same layer, it must hold that $\charac(I)\cap\charac(I')=\emptyset$. 

\looseness=-1
For a layer $\ell$, we denote the set of interactions by $\ints(\ell)$ and the timestamp of a layer by $\timest(\ell)$ (with slight abuse of notation). We focus on combinatorial storylines, as opposed to geometric storylines, meaning that our algorithm should output a (horizontal) ordering $o_L(\mathcal{S})$ of layers, and for each layer $\ell$, a (vertical) ordering $o_c(\ell)$ of the characters, and all interactions must occur in some layer.
For two interactions $I,I'$ such that $\timest(I)<\timest(I')$, let~$\ell$ and $\ell'$ be the layers of $I$ and $I'$, respectively. Then $\ell$ must be before $\ell'$ in~$o_L(\mathcal{S})$.
A character is \emph{active} in a layer if it appears in the character ordering for that layer. 
A character must be active in a contiguous range of layers including the first and last interaction it is involved in. A character is active in a layer if it appears in the character ordering for that layer.

\subparagraph*{Contributions.}
In this paper we introduce the time interval storylines model, as well as two methods to compute layer and character orderings. In~\cref{sec:thesis_algs} we introduce an algorithmic pipeline based on ILP formulations and heuristics that computes time interval storylines. We further present an ILP formulation that outputs a crossing-minimal time interval storyline in~\cref{sec:ilp}. Lastly in~\cref{sec:eval}, we experimentally evaluate our pipeline and ILP formulation.

\section{Computing combinatorial storylines}

\subsection{A pipeline heuristic}\label{sec:thesis_algs}

As the traditional storyline crossing minimization problem is a restricted version of the time interval formulation, our problem is immediately \NP-hard~\cite{KostitsynaNP0S15}. Thus, we first aim to design an efficient heuristic to generate time interval storylines, which consists of the following stages.
\begin{enumerate}[(i)]
    \item Initially, we assign each interaction to a layer,
    \item then, we compute a horizontal ordering~$o_L(\mathcal{S})$ of the layers obtained in step (i), and
    \item finally, we compute a vertical ordering~$o_c(\ell)$ of the characters for each layer~$\ell\in o_L(\mathcal{S})$.
\end{enumerate}

For step (i), the assignment is obtained using graph coloring. For each $t\in T$, we create a conflict graph $G_t=(\mathcal{I}_t,E)$ where $\mathcal{I}_t\subseteq\mathcal{I}$ and $I\in \mathcal{I}_t$ if and only if $\timest(I)=t$. Two interactions are connected by an edge if they share at least one character. Each color class then corresponds to a set of interactions which share no characters and can appear together in a layer.
We solve this problem using a straightforward ILP formulation based on variables $x_{v,c}=1$ if color $c$ is assigned to vertex $v$ and $0$ otherwise.
We can choose to limit the size of each color class \new{by adding an upper bound on the number of interactions assigned to each color}, which forces fewer interactions per layer. \new{While this allows us to limit the height of each slice, it} likely results in more layers. 

To compute a horizontal ordering of the layers in step (ii), we use a traveling salesperson (TSP) model. Concretely, for the slice corresponding to the timestamp $t$, we create a complete weighted graph $G=(\mathcal{L},E)$, where $\mathcal{L}$ corresponds to all the layers $\ell$ such that $\timest(\ell)=t$. 
For each edge $e$ between a pair of layers $\ell$ and $\ell'$ in $\mathcal{L}$, we associate a weight $w_e$, estimating the number of crossings that may occur if the two layers are consecutive as follows.

\begin{figure}
	\centering
	\includegraphics{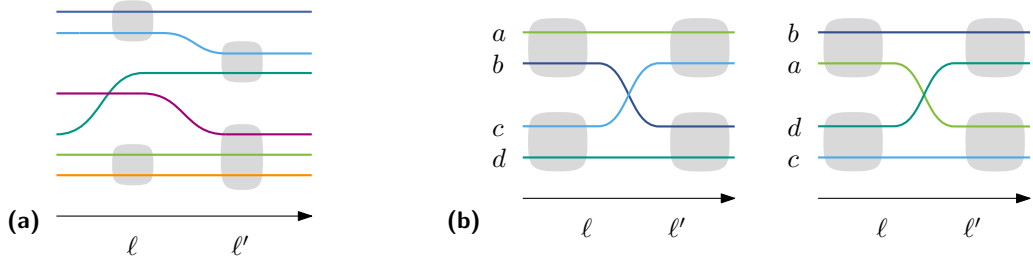}
	\caption{\textbf{\textsf{(a)}} The orange and light green characters are together in two interactions, which increases similarity between $\ell$ and $\ell'$, but the two blue characters are once together and once apart, which decreases similarity. \textbf{\textsf{(b)}} An example of an unavoidable crossing pattern.
	}
	\label{fig:pattern}
\end{figure}

Minimizing \new{the} crossings \new{of the curves representing the characters} is \NP-complete~\cite{garey83,KostitsynaNP0S15}, thus we propose two heuristics to estimate the number of crossings.
First, we propose to use set similarity measures to describe how similar the interactions in two layers $\ell$ and $\ell'$ are: If $\ell$ and $\ell'$ both have an interaction that contains the same set of characters, then no crossing should be induced by the curves corresponding to those characters, when these two layers are consecutive (see~\cref{fig:pattern}a).
Second, we consider pattern matching methods that guess how many crossings could be induced by a certain ordering of the characters. There are certain patterns of interactions between two layers for which a crossing is unavoidable (see~\cref{fig:pattern}b). We count how many of these patterns occur between each pair of layers in $G$ and set the weight of the corresponding edge to that crossing count.

\subparagraph*{Set similarity.}

We propose the use of the Rand index to evaluate how similar two layers are to one another. For the layers $\ell$ and $\ell'$, the Rand index is calculate in the following manner.
\begin{itemize}
    \item We count how many character pairs are together in an interaction in $\ell$ and $\ell'$ ($n_1$).
    \item We count how many character pairs are in different interactions in $\ell$ and in $\ell'$ ($n_2$).
    \item We count how many character pairs are in different interactions in $\ell$ and in the same interaction in $\ell'$ ($n_3$).
    \item We count how many character pairs are together in an interaction in $\ell$ and not in $\ell'$ ($n_4$).
\end{itemize}

The Rand index~\cite{rand1971objective} is then given by the value $\frac{n_1+n_2}{n_1+n_2+n_3+n_4}$. If this value is closer to one then most interactions between $\ell$ and $\ell'$ have a similar set of characters. 

\subparagraph*{Pattern matching.}

Consider the four interactions $I_1,I_2,I_3,I_4$, all having timestamp $t$. We set $\charac(I_1)={a,b}$, $\charac(I_2)={c,d}$, $\charac(I_3)={a,c}$ and $\charac(I_4)={b,d}$. Naturally these four cannot be in the same layer, so we assume $I_1$ and $I_2$ are in layer $\ell_1$ and $I_3$ and $I_4$ are in layer $\ell_2$. Then, if $\ell_1$ and $\ell_2$ are consecutive, there is necessarily a crossing induced by the curve of actor $b$ and $c$ or $a$ and $d$. We count how often this pattern occurs between each layer pair in the same slice.

\medskip
To finish step (ii), we solve the \new{path formulation of the} TSP problem on $G$ and find a horizontal ordering of the layers for each time slice. We have now obtained a traditional storyline, in which each interaction belongs to a specific layer, and all layers are totally ordered. Thus, we can solve step (iii) using the state-of-the-art crossing minimization ILP by Gronemann et al.~\cite{GronemannJLM16}.

We call the pipeline variants \Piplsim\ and \Piplpat, when using the set similarity heuristic and the pattern matching heuristic in step (ii), respectively.

\subsection{ILP formulations}\label{sec:ilp}

Crossing minimization in storylines is generally solved using ILP formulations~\cite{GronemannJLM16, DijkFFLMRSW17}. We propose two formulations to handle slices, which build on the ideas of Gronemann et al.~\cite{GronemannJLM16}.
Both formulations will give us an assignment of interactions to layers, that are already totally ordered, and an ordering of characters per layer. 
For each timestamp $t\in T$, let $\mathcal{L}_t$ be a set of $|\{I\mid \timest(I)=t\}|$ layers corresponding the number of interactions at $t$, and let $\mathcal{L}=\bigcup_{t\in T}\mathcal{L}_t$. In the first formulation we assume that a character $c$ is active in all layers between the first timestamp and last timestamp, inclusively, \new{where there exists an interaction $I$ such that $c\in \charac(I)$}.
In the second formulation we will introduce additional variables that model whether a character really needs to be active, \new{since, in fact,  character curves do not need to be active before their first interaction or after their last interaction}. 
\new{In contrast to the pipeline approach, the presented ILP formulations are able to find the crossing-minimal solution for the explored search space.}
For both formulations we also present an adaptation that allows for the minimum number of layers.

\subparagraph*{First formulation.}

Let $\mathcal{C}_\ell$ be the characters that appear in layer $\ell\in \mathcal{L}$, as discussed before.
First we introduce for each $t\in T$ the binary variables $y_{\ell,I}$ for $\ell\in \mathcal{L}_t$ and $I\in \mathcal{I}$ where $\timest(I)=t$. These should be one iff interaction $I$ is assigned to layer $\ell$.
This is realized by constraints of type~\eqref{eq:interactionassign}.
If two different interactions $I$ and $I'$ share a character they cannot be in the same layer, realized by type~\eqref{eq:interactionintersection} constraints.
\begin{align}
    \sum_{\ell\in \mathcal{L}_t}y_{\ell,I}&=1&t\in T,I\in \mathcal{I},\timest(I)=t\label{eq:interactionassign}\\
    y_{\ell,I}+y_{\ell,I'}&\le 1&\timest(I)=\timest(I')=t,\charac(I)\cap \charac(I')\ne\emptyset,\ell\in \mathcal{L}_t\label{eq:interactionintersection}
\end{align}
Next we introduce binary ordering variables $x_{\ell,c_i,c_j}$ for each layer $\ell\in \mathcal{L}$ and $c_i,c_j\in \mathcal{C}_{\ell}$ with $i<j$. Variable $x_{\ell,c_i,c_j}$ should be one iff $c_i$ comes before $c_j$ on layer $\ell$.
Standard transitivity constraints \eqref{eq:transitivity} (see e.g.~\cite{DBLP:journals/mp/GrotschelJR85a}) ensure that the binary variables induce a total order.
\begin{align}
    0\le x_{\ell,c_h,c_i}+x_{\ell,c_i,c_j}-x_{\ell,c_h,c_j}&\le 1&c_i,c_j,c_h\in \mathcal{C_\ell},i<j<h\label{eq:transitivity}
\end{align}
The crux is now to model the assignment of some interaction $I$ to some layer $\ell$, linking the $x$- and $y$-variables together.
This is done with so-called tree-constraints~\cite{GronemannJLM16}:
Let $\ell\in \mathcal{L}_t$, $I\in \mathcal{I}$ with $\timest(I)=t$ and $c_i,c_j,c_k\in \mathcal{C}_\ell$ such that $i<j$, $c_i,c_j\in \charac(I)$, and $c_k\not\in\charac(I)$.
If $i<j<k$ we add constraints \eqref{eq:tree1} and \eqref{eq:tree2}, which ensure that $c_k$ is either before or after both $c_i$ and $c_j$. 
\begin{align}
    x_{\ell,c_i,c_k}&\le x_{\ell,c_j,c_k}+\new{1-y_{\ell,I}}\label{eq:tree1}\\
    x_{\ell,c_j,c_k}&\le x_{\ell,c_i,c_k}+\new{1-y_{\ell,I}}\label{eq:tree2}
\end{align}
Similarly, if $k<i<j$ we add constraints of type~\eqref{eq:tree3} and~\eqref{eq:tree4}.
\begin{align}
    x_{\ell,c_k,c_i}&\le x_{\ell,c_k,c_j,}+\new{1-y_{\ell,I}}\label{eq:tree3}\\
    x_{\ell,c_k,c_j}&\le x_{\ell,c_k,c_i}+\new{1-y_{\ell,I}}\label{eq:tree4}
\end{align}
Lastly, if $i<k<j$ we add constraints 
 of type~\eqref{eq:tree5} and~\eqref{eq:tree6}.
\begin{align}
    x_{\ell,c_i,c_k}&\le x_{\ell,c_k,c_j}+\new{1-y_{\ell,I}}\label{eq:tree5}\\
    x_{\ell,c_k,c_j}&\le x_{\ell,c_i,c_k}+\new{1-y_{\ell,I}}\label{eq:tree6}
\end{align}
Lastly, to \new{optimize} the number of crossings we have to provide an objective function.
For this we introduce binary variables $z_{\ell,c_i,c_j}$ for all layers $\ell$ but the rightmost one and all $c_i,c_j\in\mathcal{C}_{\ell}\cap\mathcal{C}_{\ell'}$ where $\ell'$ is the adjacent layer of $\ell$ to the right. Variable $z_{\ell,c_i,c_j}$ should be one iff the character lines of $c_i$ and $c_j$ cross between layers $\ell$ and $\ell'$.
Linking variables $z_{\ell,c_i,c_j}$ is done by introducing the constraints corresponding to setting $z_{\ell,c_i,c_j}\ge x_{\ell,c_i,c_j}\new{\oplus} x_{\ell',c_i,c_j}$ where $x\oplus y$ denotes the exclusive-or relation of two binary variables $x$ and $y$. This is done with the following constraints.
\begin{align}
    z_{\ell,c_i,c_j}\ge x_{\ell,c_i,c_j}-x_{\ell',c_i,c_j}\label{eq:xor1}\\
    z_{\ell,c_i,c_j}\ge x_{\ell',c_i,c_j}-x_{\ell,c_i,c_j}\label{eq:xor2}
\end{align}
The objective is then to simply minimize $\sum z_{\ell,c_i,c_j}$. A solution to the ILP model is then transformed into a storyline realization of the input. We call this formulation \ILPone.

In the above formulation we have one layer for each interaction, which does not utilize the potential of having multiple interactions in one layer. We can, however, minimize the number of layers beforehand, using the graph coloring problem as in \cref{sec:thesis_algs}.
If we need $q$ colors for timestamp $t$, we let $\mathcal{L}_t$ only consist of $q$ layers. This can of course result in more crossings in the end. We call this adapted formulation \ILPoneml.

\subparagraph*{Second Formulation.}
In the above models \ILPone\ and \ILPoneml\ a character was contained in all layers of the first and last timestamp that contains an interaction, in which the character appears.
We also present a second ILP formulation called \ILPtwo\ that accounts for the active range of a character. This formulation contains the same variables as \ILPone\ plus the binary variables $a_{c,\ell}$ for all $c\in\mathcal{C}$ and all $\ell\in\mathcal{L}$ such that $c\in\mathcal{C}_\ell$. If $c\in C_\ell$ this does not mean that $c$ will be in layer $\ell$ when transforming a solution of the ILP to a storyline realization. This is only the case if additionally $a_{c,\ell}$ is one.
The formulation \ILPtwo\ contains the constraints for layers \eqref{eq:interactionassign} and \eqref{eq:interactionintersection}, the transitivity-constraints \eqref{eq:transitivity}, the tree-constraints \eqref{eq:tree1},\eqref{eq:tree2}, and \eqref{eq:tree3}-\eqref{eq:tree6}, and the objective function from formulation \ILPone.
Additionally, it contains the following constraints.
First, if character $c$ appears in some interaction $I$ that is present in layer $\ell$, then $c$ has to be active in $\ell$ (see \eqref{eq:activeinteraction}).
\begin{align}
    a_{c,\ell}&\ge y_{\ell,I}&I\in\mathcal{I},c\in\mathcal{C},\ell\in\mathcal{L},c\in\mathcal{C}_\ell,\timest(I)=\timest(\ell)\label{eq:activeinteraction}
\end{align}
Now for each three different layers $\ell,\ell'$, and $\ell''$ such that $\ell$ appears first, $\ell'$ appears second, and $\ell''$ appears third in the ordering of layers, let $c\in \mathcal{C}_{\ell}\cap \mathcal{C}_{\ell'}\cap \mathcal{C}_{\ell''}$. We have to ensure that if $c$ is active in $\ell$ and $\ell''$, it also has to be active in $\ell'$, done with constraints of type~\eqref{eq:activepropag}.
\begin{align}
    a_{c,\ell'}+1&\ge a_{c,\ell}+a_{c,\ell''}\label{eq:activepropag}
\end{align}
Lastly, we only have to count crossings between two character lines in two layers, if both characters are active in the corresponding layers.
Hence, let $c$ and $c'$ be two different characters and let $\ell$ and $\ell'$ be two consecutive layers such that $c,c'\in\mathcal{C}_\ell\cap\mathcal{C}_{\ell'}$.
We transform the xor-constraints \eqref{eq:xor1} and \eqref{eq:xor2} into the new constraints \eqref{eq:xornew1} and \eqref{eq:xornew2}.
Notice that none of the constraints have an effect on the $z$-variable, if at least one of the $a$-variables is zero.
\begin{align}
    z_{\ell,c_i,c_j}\ge x_{\ell,c_i,c_j}-x_{\ell',c_i,c_j}-4+a_{c,\ell}+a_{c,\ell'}+a_{c',\ell}+a_{c',\ell'}\label{eq:xornew1}\\
    z_{\ell,c_i,c_j}\ge x_{\ell',c_i,c_j}-x_{\ell,c_i,c_j}-4+a_{c,\ell}-a_{c,\ell'}+a_{c',\ell}+a_{c',\ell'}\label{eq:xornew2}
\end{align}

We can then introduce the fewest possible number of layers as for \ILPone, resulting in formulation \ILPtwoml.

\section{Experimental Evaluation}\label{sec:eval}
We performed a set of experiments to evaluate all of our presented algorithms.
In the following we describe the datasets, implementation and setup, and elaborate on the results.
\subsection{Datasets}\label{sec:appdata}
Our dataset consists of 7 instances provided by different sources. All of them come from work on storyline visualizations. General statistics on these datasets can be found in \cref{tab:datastata}.
The datasets gdea10 and gdea20 are from the work of of Gronemann et al.~\cite{GronemannJLM16} and consist of a set of publications from 1994 to 2019. These publications are the interactions and the characters are the union of all authors of these publications. The years of interactions are set as their timestamps.
Similarly, datasets ubiq1 and ubiq2 is publication data of a work by Giacomo et al.~\cite{GiacomoDLMT20} on storylines with ubiquituous characters.
Interactions, characters, and timestamp are determined as in the gdea10 and gdea20 datasets.
The last three datasets are interactions between characters of a book. These consist of the first chapter of \emph{Anna Karenina} (anna1), the first chapter of \emph{Les Misérables} (jean1), and the entirety of \emph{Huckleberry Finn} (huck). The timestamps of interactions are taken as the scenes of the book (or chapter).
\begin{table}[!t]
    \centering
    \caption{Statistics for the input data sets. The column $|\mathcal{L}|$-coloring shows the number of layers achieved by the graph coloring pipeline step, which is used for the algorithms \Piplsim, \Piplpat, \ILPoneml, and \ILPtwoml.}
    \label{tab:datastata}
    \begin{tabular}{l | >{\centering\arraybackslash}p{2cm} | >{\centering\arraybackslash}p{2cm}| >{\centering\arraybackslash}p{2cm}| >{\centering\arraybackslash}p{2cm}}
         \toprule
         & $|\mathcal{I}|$ & $|\mathcal{C}|$ & $|T|$ & $|\mathcal{L}|$-coloring\\\midrule
         gdea10 & 41 & 9 & 16 & 35 \\
         gdea20 & 100 & 19 & 17 & 47 \\
         ubiq1 & 41 & 5 & 19 & 41 \\
         ubiq2 & 45 & 5 & 18 & 38 \\
         anna1 & 58 & 41 & 34 & 53  \\
         jean1 & 95 & 40 & 65 & 88 \\
         huck & 107 & 74 & 43 & 81 \\     \bottomrule
    \end{tabular}
\end{table}

\subsection{Implementation and Setup}
All implementations were done in Python3. For the graph coloring step of the algorithms we used a simple ILP formulation. For the TSP formulation in the algorithms \Piplpat\ and \Piplsim\ we used a simple subtour elimination formulation. As the graph coloring and TSP steps of our algorithms take negligible time when compared to the crossing minimization, we do not provide more details here.

All ILP formulations in the algorithms \Piplpat\ and \Piplsim\ were solved using CPLEX 22.1 and were run on a local machine with Linux and an Intel i7-8700K 3.70GHz CPU. All formulations in the algorithms \ILPone, \ILPoneml, \ILPtwo, and \ILPtwoml\ were solved using Gurobi 9.5.1 and were run on a cluster with an AMD EPYC 7402, 2.80GHz 24-core CPU. No multithreading was used in any algorithm.

The timeout for all algorithms was set to 3600 seconds. If we run into a timeout during one of our algorithms we report the best feasible solution (in \cref{tab:crossings}) and the gap to the best lower bound found by the ILP solver (in \cref{tab:time}).

\subsection{Results}
Next, we provide the results on layer minimization, number of crossings and runtime. Examples for storylines produced by our algorithms are given in \cref{sec:examplestorylines}.
\subparagraph*{Layer minimization.}
\cref{tab:datastata} shows the number of layers that was achieved by the graph coloring pipeline step ($|\mathcal{L}|$-coloring).
The layer minimization step is applied by the algorithms \Piplsim, \Piplpat, \ILPoneml, and \ILPtwoml.
When comparing these numbers with the upper bound of the number of interactions $|\mathcal{I}|$ we can see that for some of the datasets we could reduce the number of layers significantly. There is only one dataset where we could not reduce the number of required layers.

\begin{table}[!t]
  \centering
  \caption{The number of crossings achieved by the different algorithms for all the datasets. If a number is red, then the approach timed out and we only report the best upper bound of a feasible solution given by the ILP solver.}
    \label{tab:crossings}
  \begin{tabular}{l | >{\centering\arraybackslash}p{1.6cm} | >{\centering\arraybackslash}p{1.6cm}| >{\centering\arraybackslash}p{1.6cm}| >{\centering\arraybackslash}p{1.6cm}| >{\centering\arraybackslash}p{1.6cm}| >{\centering\arraybackslash}p{1.6cm}}
    \toprule
     & \Piplsim & \Piplpat  &\ILPone &\ILPoneml &\ILPtwo & \ILPtwoml\\\midrule
     gdea10 & 6 & 6 & 7 & 7 & 6 & 6 \\
     gdea20 & 38 & 38 & \textcolor{red}{44} & \textcolor{red}{42} & \textcolor{red}{796} & \textcolor{red}{34} \\
     ubiq1 & 10 & 9 & 10 & 10 & 8 & 8 \\
     ubiq2 & 19 & 17 & 15 & 15 & 14 & 15 \\
     anna1 & 19 & 19 & 23 & 23 & 16 & 16 \\
     jean1 & 12 & 12 & \textcolor{red}{23} & \textcolor{red}{25} & \textcolor{red}{3} & \textcolor{red}{3} \\
     huck & 44 & 42 & \textcolor{red}{58} & \textcolor{red}{55} & \textcolor{red}{59} & \textcolor{red}{45} \\     
    \bottomrule
  \end{tabular}
\end{table}
\subparagraph*{Crossings.}
\cref{tab:crossings} shows the number of crossings for all algorithms and datasets. If an algorithm times out we provide the number of crossings for the best feasible solution. We can see that the ILP-formulations produce \new{fewer} crossings than the pipeline-approaches whenever they do not time out. 
Further, even if they time out, sometimes the best feasible solution is better than the solution produced by the pipeline-approaches.
It is also the case that the ILP formulations \ILPtwo\ and \ILPtwoml\ using activity-variables achieve fewer crossings than the formulations \ILPone\ and \ILPoneml\ without these additional variables. Another interesting observation is that the layer minimization does not have a negative effect on the number of crossings in most cases (see \ILPone vs.\ \ILPoneml\ and \ILPtwo\ vs. \ILPtwoml).
\new{The size of the solution space depends on the combination of number of characters, number of interactions, and number of layers, so it is hard to quantify the effect of a single input parameter onto the runtime. But generally, the increase of any of these parameters also increases the solution space and thus also the runtime of our algorithms.}

\begin{table}[!t]
  \centering
  \caption{The runtime for the different algorithms and datasets in seconds. If an approach timed out (3600 $s$), we report the relative gap between best lower and upper bound in percent given by the ILP solver in red. If the gap is 100\% then no lower bound greater than zero was found, otherwise it is calculated as $(\textsc{UB}-\textsc{LB})/\textsc{UB}$ where $\textsc{UB}$ is the upper bound and $\textsc{LB}$ is the lower bound.}
    \label{tab:time}
  \begin{tabular}{l|>{\centering\arraybackslash}p{1.6cm} | >{\centering\arraybackslash}p{1.6cm}| >{\centering\arraybackslash}p{1.6cm}| >{\centering\arraybackslash}p{1.6cm}| >{\centering\arraybackslash}p{1.6cm}| >{\centering\arraybackslash}p{1.6cm}}
\toprule  & \Piplsim & \Piplpat  &\ILPone &\ILPoneml &\ILPtwo & \ILPtwoml\\\midrule
     gdea10 & 4 & 4 & 32 & 10 & 40 & 12 \\
     gdea20 & 114 & 141 & \textcolor{red}{100\%} & \textcolor{red}{88\%} & \textcolor{red}{100\%} & \textcolor{red}{91\%} \\
     ubiq1 & 3 & 3 & 45 & 45 & 24 & 25 \\
     ubiq2 & 4 & 4 & 1039 & 176 & 1291 & 281 \\
     anna1 & 16 & 16 & 1826 & 330 & 1108 & 289 \\
     jean1 & 19 & 19 & \textcolor{red}{95\%} & \textcolor{red}{92\%} & \textcolor{red}{100\%} & \textcolor{red}{33\%} \\
     huck & 136 & 127 & \textcolor{red}{84\%} & \textcolor{red}{78\%} & \textcolor{red}{95\%} & \textcolor{red}{73\%}  \\    
    \bottomrule
  \end{tabular}
\end{table}
\subparagraph*{Runtime.}
\cref{tab:time} reports the runtimes of the algorithms. If an algorithm times out then we report the gap between the best feasible solution and best lower bound found by the solver. As expected, the runtimes of the ILP-algorithms are much higher than the pipeline-algorithms.
Further, minimizing the number of layers seems to positively affect the runtime of the ILP-approaches. This is due to the fact that having less layers also reduces the search space of the ILP-formulations. The optimality gaps of the ILP-approaches on instances that timed out are quite high, so we do not expect that the solvers would find an optimal solution for these instances in a feasible time. Additional experiments even showed that most of these instances could not be solved by our ILP-formulations optimally even after one week.

\subsection{Example Storylines}\label{sec:examplestorylines}
\begin{figure}
  \centering
  \begin{tabular}[b]{l}
    {\centering
    \includegraphics[width=0.9\linewidth]{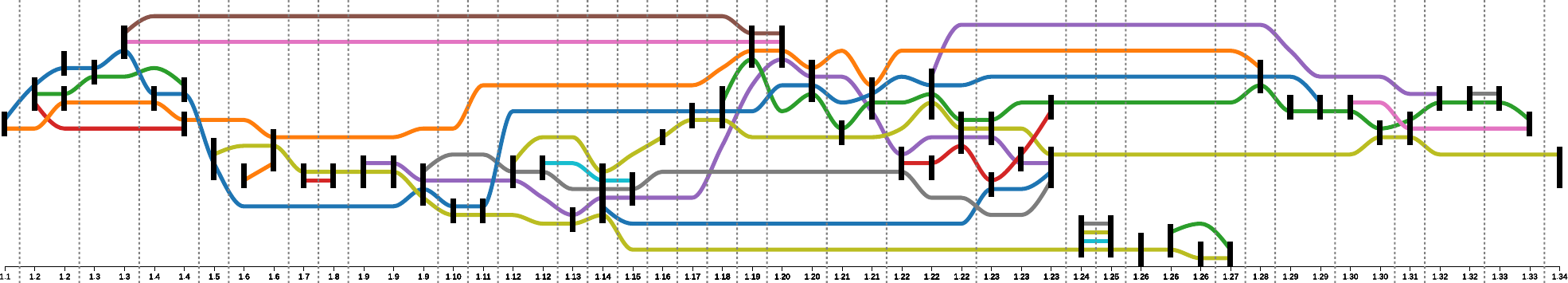} }\\
    \vspace{\abovecaptionskip}
    \small \textbf{\textsf{(a)}} Algorithm \Piplsim: 53 layers and 19 crossings.
  \end{tabular}
  \medskip
  \begin{tabular}[b]{l}
    {\centering
    \includegraphics[width=0.9\linewidth]{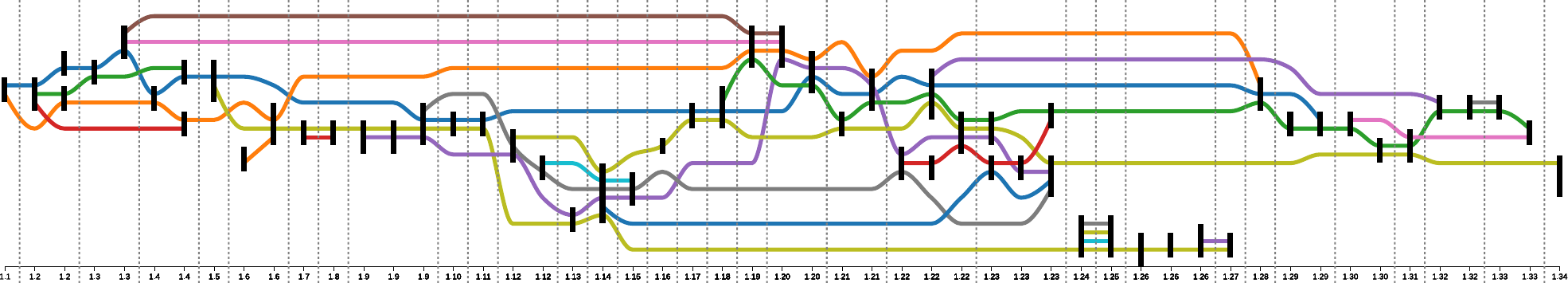} }\\
    \vspace{\abovecaptionskip}
    \small \textbf{\textsf{(b)}} Algorithm \Piplpat: 53 layers and 19 crossings.
  \end{tabular}
  \medskip
  \begin{tabular}[b]{l}
    {\centering
    \includegraphics[width=0.9\linewidth]{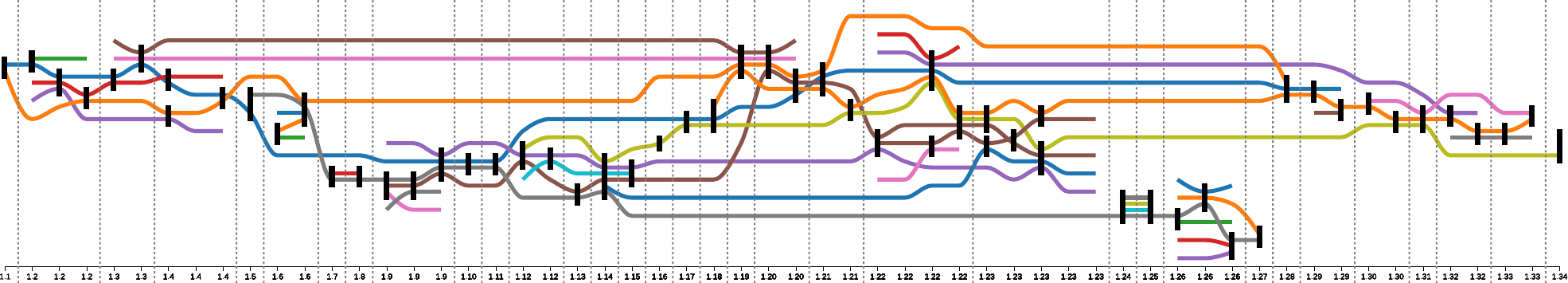} }\\
    \vspace{\abovecaptionskip}
    \small \textbf{\textsf{(a)}} Algorithm \ILPone: 53 layers and 23 crossings.
  \end{tabular}
  \medskip
  \begin{tabular}[b]{l}
    {\centering
    \includegraphics[width=0.9\linewidth]{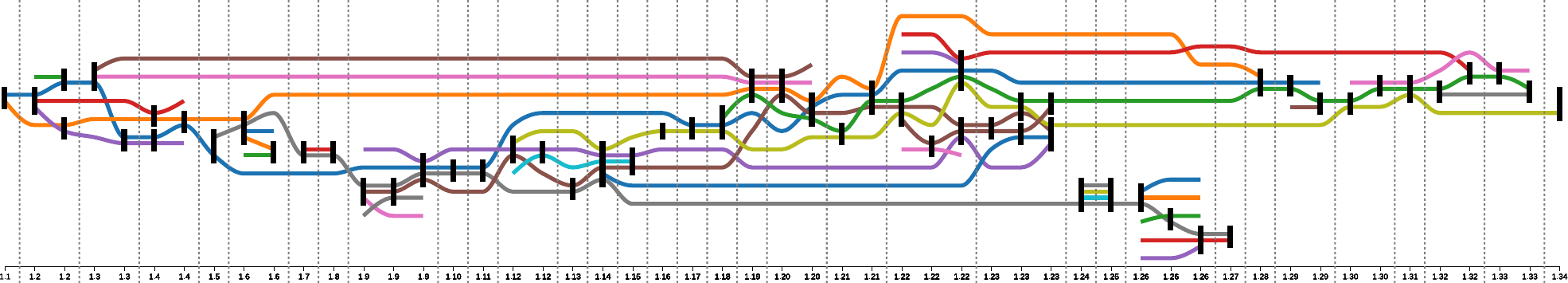} }\\
    \vspace{\abovecaptionskip}
    \small \textbf{\textsf{(c)}} Algorithm \ILPoneml: 53 layers and 23 crossings.
  \end{tabular}
  \medskip
  \begin{tabular}[b]{l}
    {\centering
    \includegraphics[width=0.9\linewidth]{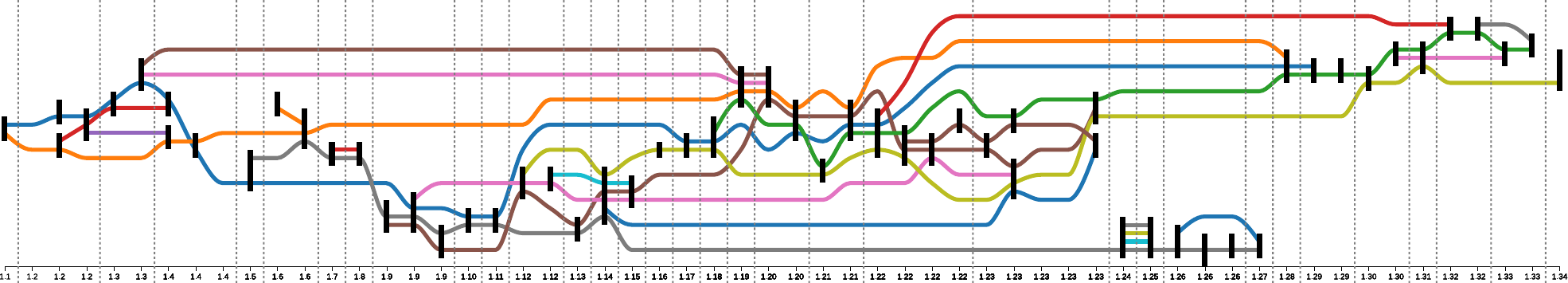} }\\
    \vspace{\abovecaptionskip}
    \small \textbf{\textsf{(d)}} Algorithm \ILPtwo: 58 layers and 16 crossings.
  \end{tabular}
  \medskip
  \begin{tabular}[b]{l}
    {\centering
    \includegraphics[width=0.9\linewidth]{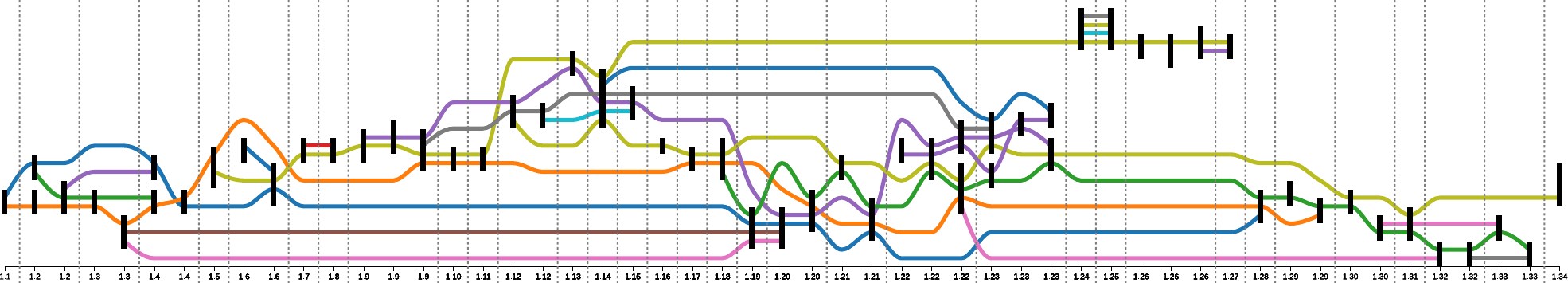} }\\
    \vspace{\abovecaptionskip}
    \small \textbf{\textsf{(e)}} Algorithm \ILPtwoml: 53 layers and 16 crossings.
  \end{tabular}
  \caption{The storylines for dataset anna1 and the algorithms \Piplsim, \Piplpat, \ILPone, \ILPoneml, \ILPtwo, and \ILPtwoml. The $x$-axes are labeled by the scenes of the book, which are separated by dashed gray lines and correspond to the time intervals. Interactions are visualized with black vertical bars and correspond to the characters in the book interacting with each other shown as $x$-monotone curves.}
  \label{fig:remainingexamples}
\end{figure}

\Cref{fig:remainingexamples} show storylines created by the different algorithms and the anna1 dataset. We have applied a simple wiggle-height minimization post-processing algorithm to the purely combinatorial output of our algorithms. This post-processing algorithm assigns actual $x$- and $y$-coordinates to the character lines, and is based on a similar approach that was applied to drawing directed graphs \cite{DBLP:journals/tse/GansnerKNV93}. Notice that in the outputs for \ILPone\ and \ILPoneml\ character lines sometimes start before their first interaction and end before their last interaction. This is never the case for all other algorithms. But those algorithms have the problem, that character curves are often very short, making it hard to follow these curves. Further, some characters only appear for one layer (see the last layer of all shown storylines). In follow-up visualizations we should add some offset to these curves so they are visible. \new{We do not dare to give a qualitative judgement about which of the visualizations looks best, but generally the visual clarity of the visualizations is increased due to having few crossings.}

\section{Conclusion}
We introduced storylines with time intervals, which capture many real-world datasets, such as authorship networks with multiple papers per year. We also provide two methods to compute the vertical and horizontal orderings in these storylines. Preliminary experiments show that these storylines allow for more effective width-minimization, by assigning multiple interactions to one (vertical) layer, and more effective crossing-minimization, by exploiting different horizontal orderings of interactions. 
Further research directions include optimizing our current methods and extending the model further, for example with overlapping slices.

\bibliography{eurocg23bib}

\newpage
\appendix

\end{document}